\documentclass[showpacs,aps,graphicx,twocolumn]{revtex4}%and
\usepackage{graphicx}
\begin{document}

\title{Distillation of genuine mixed state for quantum communications}

\author{Yu-Bo Sheng,$^{1,2}$\footnote{Email address:
shengyb@njupt.edu.cn}   Lan Zhou$^{2,3}$, Xing-Fu Wang$^{3}$ }
\address{$^1$Institute of Signal Processing  Transmission, Nanjing
University of Posts and Telecommunications, Nanjing, 210003,  China\\
 $^2$Key Lab of Broadband Wireless Communication and Sensor Network
 Technology,
 Nanjing University of Posts and Telecommunications, Ministry of
 Education, Nanjing, 210003,
 China\\
 $^3$College of Mathematics \& Physics, Nanjing University of Posts and Telecommunications, Nanjing,
210003, China\\}

\begin{abstract}
We present a practical  entanglement distillation protocol (EDP) with linear optics for a genuine mixed state.  Each components
of the genuine mixed state is  a pure less-entangled state.
 After successfully performing this EDP, we can  obtain a high quality entangled mixed state. Our EDP can work for both ideal entanglement sources and current  available spontaneous parametric
down-conversion (SPDC) sources, which makes it feasible in current experimental technology.  Moreover, by using the SPDC source, we can obtain a higher fidelity. This protocol can also be used to distill the multi-partite entangled systems. All the features make it practical and useful in
current quantum communications.
\end{abstract}
 \maketitle

\section{Introduction}
Entanglement plays the most important role in quantum information processing \cite{book,rmp}.
 The well known quantum teleportation \cite{teleportation,cteleportation}, quantum key distribution \cite{Ekert91,QKDdeng1,QKDdeng2},
  quantum secret sharing \cite{QSS1,QSS2,QSS3}, quantum
 secure direct communication (QSDC) \cite{QSDC1,QSDC2,QSDC3}, and quantum dense coding \cite{densecoding1,densecoding2},
 all resort to the entanglement
 for setting up the quantum channel between long-distance locations.
Unfortunately, the noise from environment will decrease the quality of the quantum
 channel and cause the entanglement decoherence.  Generally speaking, the decoherence are regarded as
 two different ways. The first one is that
 the maximally entangled state $|\phi^{+}\rangle$ will become a mixed state
 say $\rho_{\circ}=F|\phi^{+}\rangle\langle\phi^{+}|+(1-F)|\psi^{+}\rangle\langle\psi^{+}|$. Here $|\phi^{+}\rangle=\frac{1}{\sqrt{2}}(|00\rangle+|11\rangle)$,
 and $|\psi^{+}\rangle=\frac{1}{\sqrt{2}}(|01\rangle+|10\rangle)$. The second one is that
the original state $|\phi^{+}\rangle$ will become a pure less-entangled state say $|\phi^{+}\rangle'=\alpha|00\rangle+\beta|11\rangle$, with
 $|\alpha|^{2}+|\beta|^{2}=1$.

If the maximally entangled state is polluted, it will make the fidelity
of the quantum teleportation degrade \cite{teleportation,cteleportation}. It will also make the quantum cryptography
protocol insecure. Entanglement purification is a method by which the parties can extract a smaller
number of highly entangled pairs from a large number of low quality mixed entangled pairs \cite{C.H.Bennett1,D. Deutsch,M. Murao,Yong,Pan1,Simon,Pan2,sangouard,sangouard2,shengpra1,shengpra2,shengpra3,wangc1,wangc2,wangc3,wangc4,lixhepp,dengonestep1,dengonestep2,ren2}.
Entanglement concentration is used to recover the pure less-entangled pairs into the maximally entangled pairs \cite{C.H.Bennett2,swapping1,swapping2,Yamamoto1,Yamamoto2,zhao1,zhao2,wangxb,shengpra4,shengqic,shengsinglephotonconcentration,dengsingle,
shengwstateconcentration,ren1,shenghyb}. Both the two methods are based on the local operations and classical
communications (LOCC). The early works of entanglement purification are based on the controlled-not (CNOT) gates or similar logic operations \cite{C.H.Bennett1,D. Deutsch,M. Murao,Yong}.
However, there is no implementation of the perfect CNOT gates that could be adopted
for entanglement purification in the context of long-distance quantum communication. In 2001, based on the
linear optical elements, Pan \emph{et al.} proposed an entanglement purification protocol (EPP) and subsequently realized it
in experiment \cite{Pan1,Pan2}. We call it PBS-purification protocol. There are some other EPPs, such as the EPP based on the cross-Kerr nonlinearity \cite{shengpra1}, single-photon entanglement purification \cite{sangouard,sangouard2}, the deterministic EPP \cite{shengpra2}, one-step EPP \cite{shengpra3,lixhepp}, and so on.
On the other hand, in the area of entanglement concentration, Bennett \emph{et al.} proposed an entanglement concentration protocol (ECP) based on the collective measurement, which
is called the Schimidit projection method in 1996 \cite{C.H.Bennett2}. The ECPs using   entanglement swapping and unitary transformation are proposed \cite{swapping1,swapping2}. Zhao \emph{et al.} and Yamamoto\emph{ et al.} simplified the Schimidit projection method
and proposed two similar ECPs using optical elements, independently \cite{Yamamoto1,zhao1,Yamamoto2,zhao2}. Here we call them PBS-concentration protocols.

Actually,  in a practical noisy environment,  the entanglement does not suffer from the noise in a unique way.
Both the two ways of decoherence described above are existed simultaneously. In this way, the maximally entangled state will essentially become a more general form as
\begin{eqnarray}
\rho=F|\phi^{+}\rangle'\langle\phi^{+}|'+(1-F)|\psi^{+}\rangle'\langle\psi^{+}|'.\label{general}
\end{eqnarray}
Here $|\psi^{+}\rangle'=\alpha|01\rangle+\beta|10\rangle$. Unfortunately, existed
EPPs and ECPs described above are not suitable for distilling the ensembles of  $\rho$
into a high quality mixed state of the form of $\rho_{\circ}$ with a new fidelity $F'>F$.
Current EPPs are focused on the special mixed state $\rho_{\circ}$ and the ECPs are focused on the pure
less-entangled states. Both the EPPs and ECPs cannot completely solve the problem of the decoherence, respectively.

In this paper, we will present a realistic way to distill the general mixed state as shown in Eq. (\ref{general}). Our protocol is based on the
linear optics and the  post-selection principle.
 Interestingly, it contains both the functions of conventional
  entanglement purification and concentration. That is, after successfully performing this protocol, we can also obtain a similar higher fidelity $F'=\frac{F^{2}}{F^{2}+(1-F)^{2}}$ like PBS-purification protocol, but with a different
success probability. Each terms of the mixed state is a maximally entangled state.
 One obvious advantage is that for the bit-flip error correction, the tasks of both purification and concentration can be achieved  simultaneously in one step. Meanwhile, our protocol is also suitable for the case of
the multipartite entangled systems. Moreover, we also discuss the realization of this protocol in current  spontaneous parametric
down-conversion (SPDC) sources.

  This paper is organized as follows:
in Sec. II, we first briefly describe this entanglement distillation protocol (EDP) for a bit-flip error.  In Sec. III, we explain the
EDP for a phase-flip error. Different from the bit-flip error, we cannot distill it directly.
We should first perform the concentration and  then transform the phase-flip error into a bit-flip error, which can be
purified  in a next round. In Sec. IV, we analyze the EDP for  the multi-photon mixed ensembles. In Sec. V, we discuss
the experimental realization with currently available SPDC source.
In Sec. VI, we  present a discussion and conclusion.

\section{Distillation of the mixed state with a Bit-flip error}
From Fig. 1, suppose that the mixed state emitted from  sources S$_{1}$ and
S$_{2}$ can be written as
\begin{eqnarray}
\rho_{ab}=F|\Phi^{+}\rangle_{ab}\langle\Phi^{+}|+(1-F)|\Psi^{+}\rangle_{ab}\langle\Psi^{+}|.\label{mixed1}
\end{eqnarray}
Here $|\Phi^{+}\rangle_{ab}$ and $|\Psi^{+}\rangle_{ab}$ are both the pure less-entangled state of the form
\begin{eqnarray}
|\Phi^{+}\rangle_{ab}=\alpha|H\rangle_{a}|H\rangle_{b}+\beta|V\rangle_{a}|V\rangle_{b},
\end{eqnarray}
and
\begin{eqnarray}
|\Psi^{+}\rangle_{ab}=\alpha|H\rangle_{a}|V\rangle_{b}+\beta|V\rangle_{a}|H\rangle_{b},
\end{eqnarray}
with $|\alpha|^{2}+|\beta|^{2}=1$. $|H\rangle$ and $|V\rangle$ represent the horizonal and vertical
polarization of the photons, respectively.
\begin{figure}[!h]%[tpb]
\begin{center}
\includegraphics[width=9cm,angle=0]{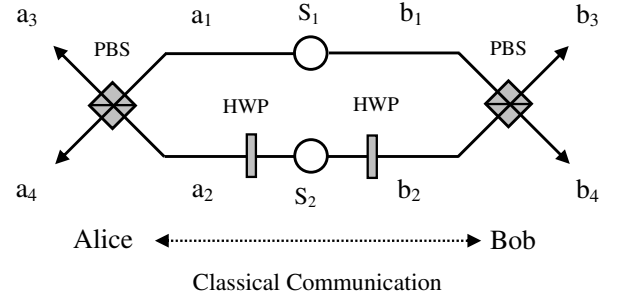}
\caption{A schematic drawing of our EDP. It is rather analogous to that of PBS-purification protocol \cite{Pan1}. We add
two half-wave plates (HWPs) in the $a_{2}$ and $b_{2}$ spatial modes. The entanglement sources $S_{1}$ and S$_{2}$
create one pair of genuine mixed state described in Eq. (\ref{mixed1}). After the four photons passing through the PBSs at the same time, by
selecting the four-mode cases, one can obtain the higher quality mixed state, where each item is also the maximally entangled state.}
\end{center}
\end{figure}

Before the photons reaching  the locations of Alice and Bob,  they first make a bit-flip operation on  the photons in the spatial modes $a_{2}$ and $b_{2}$, using the half-wave plate (HWP).
Then the state can be written as
\begin{eqnarray}
\rho_{a_{2}b_{2}}=F|\Phi^{+}\rangle^{\perp}_{a_{2}b_{2}}\langle\Phi^{+}|^{\perp}+(1-F)|\Psi^{+}\rangle^{\perp}_{a_{2}b_{2}}\langle\Psi^{+}|^{\perp}.\label{mixed2}
\end{eqnarray}
Here
\begin{eqnarray}
|\Phi^{+}\rangle^{\perp}_{a_{2}b_{2}}=\alpha|V\rangle_{a_{2}}|V\rangle_{b_{2}}+\beta|H\rangle_{a_{2}}|H\rangle_{b_{2}},
\end{eqnarray}
\begin{eqnarray}
|\Psi^{+}\rangle^{\perp}_{a_{2}b_{2}}=\alpha|V\rangle_{a_{2}}|H\rangle_{b_{2}}+\beta|H\rangle_{a_{2}}|V\rangle_{b_{2}}.
\end{eqnarray}

Our proposed protocol is rather analogous to that of Ref. \cite{Pan1}. The parties also
proceed by operating on two pairs of photons at the same time. From Fig. 1, Alice and Bob both let
his or her photons pass through the PBSs. Our distillation step is to choose the cases where there is exactly one photon in each of the four spatial output modes.
It is so called 'four-mode cases' \cite{Pan1}. From Eq. (\ref{mixed1}) and Eq. (\ref{mixed2}), the original state
of the two pairs can be seen as a probabilistic mixture of four pure states: with a probability of $F^{2}$, pairs
in the spatial modes $a_{1}b_{1}$ and $a_{2}b_{2}$ are in the state $|\Phi^{+}\rangle_{a_{1}b_{1}}\otimes|\Phi^{+}\rangle^{\perp}_{a_{2}b_{2}}$,
with equal probabilities of $F(1-F)$ in the states $|\Phi^{+}\rangle_{a_{1}b_{1}}\otimes|\Psi^{+}\rangle^{\perp}_{a_{2}b_{2}}$ and $|\Psi^{+}\rangle_{a_{1}b_{1}}\otimes|\Phi^{+}\rangle^{\perp}_{a_{2}b_{2}}$, and with a probability of $(1-F)^{2}$, they are in $|\Psi^{+}\rangle_{a_{1}b_{1}}\otimes|\Psi^{+}\rangle^{\perp}_{a_{2}b_{2}}$.

We first consider the cross-combinations $|\Phi^{+}\rangle_{a_{1}b_{1}}\otimes|\Psi^{+}\rangle^{\perp}_{a_{2}b_{2}}$ and $|\Psi^{+}\rangle_{a_{1}b_{1}}\otimes|\Phi^{+}\rangle^{\perp}_{a_{2}b_{2}}$ with the same probability of $F(1-F)$. Interestingly, they never lead
to four-mode cases. We take $|\Phi^{+}\rangle_{a_{1}b_{1}}\otimes|\Psi^{+}\rangle^{\perp}_{a_{2}b_{2}}$ as an example.
\begin{eqnarray}
&&|\Phi^{+}\rangle_{a_{1}b_{1}}\otimes|\Psi^{+}\rangle^{\perp}_{a_{2}b_{2}}
=(\alpha|H\rangle_{a_{1}}|H\rangle_{b_{1}}+\beta|V\rangle_{a_{1}}|V\rangle_{b_{1}})\nonumber\\
&\otimes&(\alpha|V\rangle_{a_{2}}|H\rangle_{b_{2}}+\beta|H\rangle_{a_{2}}|V\rangle_{b_{2}})\nonumber\\
&=&\alpha^{2}|H\rangle_{a_{1}}|V\rangle_{a_{2}}|H\rangle_{b_{1}}|H\rangle_{b_{2}}+\alpha\beta(|V\rangle_{a_{1}}|V\rangle_{a_{2}}|V\rangle_{b_{1}}|H\rangle_{b_{2}}\nonumber\\
&+&|H\rangle_{a_{1}}|H\rangle_{a_{2}}|H\rangle_{b_{1}}|V\rangle_{b_{2}})+\beta^{2}|V\rangle_{a_{1}}|H\rangle_{a_{2}}|V\rangle_{b_{1}}|V\rangle_{b_{2}}.\label{corss1}\nonumber\\
\end{eqnarray}
From Eq. (\ref{corss1}), it is obvious that items $|H\rangle_{a_{1}}|V\rangle_{a_{2}}|H\rangle_{b_{1}}|H\rangle_{b_{2}}$ and $|V\rangle_{a_{1}}|H\rangle_{a_{2}}|V\rangle_{b_{1}}|V\rangle_{b_{2}}$ will lead the two
photons in Alice's location  in the same output mode, and items $|V\rangle_{a_{1}}|V\rangle_{a_{2}}|V\rangle_{b_{1}}|H\rangle_{b_{2}}$
and $|H\rangle_{a_{1}}|H\rangle_{a_{2}}|H\rangle_{b_{1}}|V\rangle_{b_{2}}$ will lead the two photons in Bob's location  in the same output mode.
Thus, by selecting only the four-mode cases, they can eliminate the contribution of the two cross-combination items.

The remained items $|\Phi^{+}\rangle_{a_{1}b_{1}}\otimes|\Phi^{+}\rangle^{\perp}_{a_{2}b_{2}}$ and $|\Psi^{+}\rangle_{a_{1}b_{1}}\otimes|\Psi^{+}\rangle^{\perp}_{a_{2}b_{2}}$ will both lead the four-mode cases. For example,
\begin{eqnarray}
&&|\Phi^{+}\rangle_{a_{1}b_{1}}\otimes|\Phi^{+}\rangle^{\perp}_{a_{2}b_{2}}=(\alpha|H\rangle_{a_{1}}|H\rangle_{b_{1}}+\beta|V\rangle_{a_{1}}|V\rangle_{b_{1}})\nonumber\\
&\otimes&(\alpha|V\rangle_{a_{2}}|V\rangle_{b_{2}}+\beta|H\rangle_{a_{2}}|H\rangle_{b_{2}})\nonumber\\
&=&\alpha^{2}|H\rangle_{a_{1}}|V\rangle_{a_{2}}|H\rangle_{b_{1}}|V\rangle_{b_{2}}+\beta^{2}|V\rangle_{a_{1}}|H\rangle_{a_{2}}|V\rangle_{b_{1}}|H\rangle_{b_{2}}\nonumber\\
&+&\alpha\beta(|H\rangle_{a_{1}}|H\rangle_{a_{2}}|H\rangle_{b_{1}}|H\rangle_{b_{2}}+|V\rangle_{a_{1}}|V\rangle_{a_{2}}|V\rangle_{b_{1}}|V\rangle_{b_{2}}).\label{remain1}\nonumber\\
\end{eqnarray}
From above equation, it is obvious that items $|H\rangle_{a_{1}}|H\rangle_{a_{2}}|H\rangle_{b_{1}}|H\rangle_{b_{2}}$ and $|V\rangle_{a_{1}}|V\rangle_{a_{2}}|V\rangle_{b_{1}}|V\rangle_{b_{2}}$ will lead the four-mode cases, with the probability of $2|\alpha\beta|^{2}F^{2}$.

On the other hand, item $|\Psi^{+}\rangle_{a_{1}b_{1}}\otimes|\Psi^{+}\rangle^{\perp}_{a_{2}b_{2}}$ can also lead the four-mode cases as
\begin{eqnarray}
&&|\Psi^{+}\rangle_{a_{1}b_{1}}\otimes|\Psi^{+}\rangle^{\perp}_{a_{2}b_{2}}=(\alpha|H\rangle_{a_{1}}|V\rangle_{b_{1}}+\beta|V\rangle_{a_{1}}|H\rangle_{b_{1}})\nonumber\\
&\otimes&(\alpha|V\rangle_{a_{2}}|H\rangle_{b_{2}}+\beta|H\rangle_{a_{2}}|V\rangle_{b_{2}})\nonumber\\
&=&\alpha^{2}|H\rangle_{a_{1}}|V\rangle_{a_{2}}|V\rangle_{b_{1}}|H\rangle_{b_{2}}+\beta^{2}|V\rangle_{a_{1}}|H\rangle_{a_{2}}|H\rangle_{b_{1}}|V\rangle_{b_{2}}\nonumber\\
&+&\alpha\beta(|H\rangle_{a_{1}}|H\rangle_{a_{2}}|V\rangle_{b_{1}}|V\rangle_{b_{2}}+|V\rangle_{a_{1}}|V\rangle_{a_{2}}|H\rangle_{b_{1}}|H\rangle_{b_{2}}).\label{remain2}\nonumber\\
\end{eqnarray}

Therefore, by selecting the four-mode cases, they can obtain the state
\begin{eqnarray}
|\Phi^{+}_{1}\rangle_{a_{3}b_{3}a_{4}b_{4}}=\frac{1}{\sqrt{2}}(|H\rangle_{a_{3}}|H\rangle_{b_{3}}|H\rangle_{a_{4}}|H\rangle_{b_{4}}\nonumber\\
+|V\rangle_{a_{3}}|V\rangle_{b_{3}}|V\rangle_{a_{4}}|V\rangle_{b_{4}}),
\end{eqnarray}
with the probability of $2|\alpha\beta|^{2}F^{2}$,
and they can obtain the state
\begin{eqnarray}
|\Psi^{+}_{1}\rangle_{a_{3}b_{3}a_{4}b_{4}}=\frac{1}{\sqrt{2}}(|H\rangle_{a_{3}}|V\rangle_{b_{3}}|H\rangle_{a_{4}}|V\rangle_{b_{4}}\nonumber\\
+|V\rangle_{a_{3}}|H\rangle_{b_{3}}|V\rangle_{a_{4}}|H\rangle_{b_{4}}),
\end{eqnarray}
with the probability of $2|\alpha\beta|^{2}(1-F)^{2}$.
Finally, by measuring the photons in $a_{4}b_{4}$ modes in the $|\pm\rangle=\frac{1}{\sqrt{2}}(|H\rangle\pm|V\rangle)$, if the measurement result is $|++\rangle$ or
$|--\rangle$, they will get
\begin{eqnarray}
|\Phi^{+}\rangle_{a_{3}b_{3}}=\frac{1}{\sqrt{2}}(|H\rangle_{a_{3}}|H\rangle_{b_{3}}
+|V\rangle_{a_{3}}|V\rangle_{b_{3}}),
\end{eqnarray}
with the fidelity of
\begin{eqnarray}
F'&=&\frac{2|\alpha\beta|^{2}F^{2}}{2|\alpha\beta|^{2}F^{2}+2|\alpha\beta|^{2}(1-F)^{2}}\nonumber\\
&=&\frac{F^{2}}{F^{2}+(1-F)^{2}}.
\end{eqnarray}
They can also get
\begin{eqnarray}
|\Phi^{-}\rangle_{a_{3}b_{3}}=\frac{1}{\sqrt{2}}(|H\rangle_{a_{3}}|H\rangle_{b_{3}}
-|V\rangle_{a_{3}}|V\rangle_{b_{3}}),
\end{eqnarray}
with the same probability of $F'$. In order to get the $|\Phi^{+}\rangle_{a_{3}b_{3}}$, one of the parties need to perform
a phase flip operation on his or her photon. The total success probability is
\begin{eqnarray}
P=2|\alpha\beta|^{2}[(1-F)^{2}+F^{2}].
\end{eqnarray}

We have briefly explained this distillation protocol for the case of a bit-flip error. It is interesting to compare this protocol with
PBS-purification protocol.  The same points of both protocols are the four-mode cases, and after performing both protocols, they will get the same
 high quality mixed state. But in this protocol, the two pairs of  mixed states
are different because the second pair  in spatial modes $a_{2}b_{2}$ should be flipped first.  This is essentially  correspond to
the traditional EPPs like Ref. \cite{zhao1}.  Another difference is that the success probability in Ref. \cite{Pan1} is $F^{2}+(1-F)^{2}$, and here it is
$2|\alpha\beta|^{2}[(1-F)^{2}+F^{2}]$. We show that if neither of items in the mixed state is a maximally one, it will decrease the final
success probability.

\section{Distillation of a mixed state with a Phase-flip error}
In above section, we have described the EDP for the bit-flip error. They can obtain the maximally  entangled state
with the fidelity of $F'$. Certainly, during the transmission, there exists another kind of error say phase-flip error. The relative phase between
the photons in different spatial modes is sensitive to path-length instabilities, which has to be kept constant within
a fraction of the photon¡¯s wavelength.  It is
  usually caused by the fiber length dispersion or atmospheric fluctuation in a free-space transmission \cite{repeater1}.
  It has become an inherent drawback in quantum repeaters which
  is severe enough to make long-distance quantum communication extremely
  difficult \cite{repeater3,repeater2}.

A mixed state with a phase-flip error can be  described as
\begin{eqnarray}
\rho'_{ab}=F|\Phi^{+}\rangle_{ab}\langle\Phi^{+}|+(1-F)|\Phi^{-}\rangle_{ab}\langle\Phi^{-}|.\label{mixed3}
\end{eqnarray}
Here
\begin{eqnarray}
|\Phi^{-}\rangle_{ab}=\alpha|H\rangle_{a}|H\rangle_{b}-\beta|V\rangle_{a}|V\rangle_{b}.
\end{eqnarray}

To distill the state $\rho'_{ab}$, we should divide it into two steps. The principle is also shown in
Fig 1. The source S$_{1}$ and S$_{2}$ both emit a less-entangled pair of the form of Eq. (\ref{mixed3}).
The pair in spatial mode $a_{2}b_{2}$ first makes a bit-flip and becomes
\begin{eqnarray}
\rho'_{a_{2}b_{2}}=F|\Phi^{+}\rangle^{\perp}_{a_{2}b_{2}}\langle\Phi^{+}|^{\perp}+(1-F)|\Phi^{-}\rangle^{\perp}_{a_{2}b_{2}}\langle\Phi^{-}|^{\perp}.\label{mixed4}
\end{eqnarray}
Here
\begin{eqnarray}
|\Phi^{-}\rangle^{\perp}_{a_{2}b_{2}}=\alpha|V\rangle_{a_{2}}|V\rangle_{b_{2}}-\beta|H\rangle_{a_{2}}|H\rangle_{b_{2}}.
\end{eqnarray}
The original state $\rho'_{a_{1}b_{1}}\otimes\rho'_{a_{2}b_{2}}$ can also be written as a probabilistic mixture
of four pure states: with a probability of $F^{2}$, they are in the state
 $|\Phi^{+}\rangle_{a_{1}b_{1}}\otimes|\Phi^{+}\rangle^{\perp}_{a_{2}b_{2}}$, with equal probabilities of $F(1-F)$ in the
 states $|\Phi^{+}\rangle_{a_{1}b_{1}}\otimes|\Phi^{-}\rangle^{\perp}_{a_{2}b_{2}}$ and $|\Phi^{-}\rangle_{a_{1}b_{1}}\otimes|\Phi^{+}\rangle^{\perp}_{a_{2}b_{2}}$,
 and with a probability of $(1-F)^{2}$ in $|\Phi^{-}\rangle_{a_{1}b_{1}}\otimes|\Phi^{-}\rangle^{\perp}_{a_{2}b_{2}}$.
 Interestingly, in these cases, all items can lead the four-mode cases. For example,
 \begin{eqnarray}
 &&|\Phi^{+}\rangle_{a_{1}b_{1}}\otimes|\Phi^{+}\rangle^{\perp}_{a_{2}b_{2}}
 =(\alpha|H\rangle_{a_{1}}|H\rangle_{b_{1}}+\beta|V\rangle_{a_{1}}|V\rangle_{b_{1}})\nonumber\\
 &&\otimes(\alpha|V\rangle_{a_{2}}|V\rangle_{b_{2}}+\beta|H\rangle_{a_{2}}|H\rangle_{b_{2}})\nonumber\\
 &&=\alpha^{2}|H\rangle_{a_{1}}|V\rangle_{a_{2}}|H\rangle_{b_{1}}|V\rangle_{b_{2}}
 +\beta^{2}|V\rangle_{a_{1}}|H\rangle_{a_{2}}|V\rangle_{b_{1}}|H\rangle_{b_{2}}\nonumber\\
 &&+\alpha\beta(|H\rangle_{a_{1}}|H\rangle_{a_{2}}|H\rangle_{b_{1}}|H\rangle_{b_{2}}
 +|V\rangle_{a_{1}}|V\rangle_{a_{2}}|V\rangle_{b_{1}}|V\rangle_{b_{2}}).\label{phase1}\nonumber\\
\end{eqnarray}
The cross-combination $|\Phi^{+}\rangle_{a_{1}b_{1}}\otimes|\Phi^{-}\rangle^{\perp}_{a_{2}b_{2}}$ can be written as
 \begin{eqnarray}
 &&|\Phi^{+}\rangle_{a_{1}b_{1}}\otimes|\Phi^{-}\rangle^{\perp}_{a_{2}b_{2}}
 =(\alpha|H\rangle_{a_{1}}|H\rangle_{b_{1}}+\beta|V\rangle_{a_{1}}|V\rangle_{b_{1}})\nonumber\\
&&\otimes(\alpha|V\rangle_{a_{2}}|V\rangle_{b_{2}}-\beta|H\rangle_{a_{2}}|H\rangle_{b_{2}})\nonumber\\
 &&=\alpha^{2}|H\rangle_{a_{1}}|V\rangle_{a_{2}}|H\rangle_{b_{1}}|V\rangle_{b_{2}}
 -\beta^{2}|V\rangle_{a_{1}}|H\rangle_{a_{2}}|V\rangle_{b_{1}}|H\rangle_{b_{2}}\nonumber\\
 &&+\alpha\beta(-|H\rangle_{a_{1}}|H\rangle_{a_{2}}|H\rangle_{b_{1}}|H\rangle_{b_{2}}
 +|V\rangle_{a_{1}}|V\rangle_{a_{2}}|V\rangle_{b_{1}}|V\rangle_{b_{2}}).\label{phase2}\nonumber\\
\end{eqnarray}
If they choose the four-mode cases, they will get
\begin{eqnarray}
|\Phi^{+}_{1}\rangle_{a_{3}b_{3}a_{4}b_{4}}=\frac{1}{\sqrt{2}}(|H\rangle_{a_{3}}|H\rangle_{b_{3}}|H\rangle_{a_{4}}|H\rangle_{b_{4}}\nonumber\\
+|V\rangle_{a_{3}}|V\rangle_{b_{3}}|V\rangle_{a_{4}}|V\rangle_{b_{4}}),
\end{eqnarray}
with the probability of $2|\alpha\beta|^{2}[F^{2}+(1-F)^{2}]$,
and get  the state
\begin{eqnarray}
|\Phi^{-}_{1}\rangle_{a_{3}b_{3}a_{4}b_{4}}=\frac{1}{\sqrt{2}}(|H\rangle_{a_{3}}|H\rangle_{b_{3}}|H\rangle_{a_{4}}|H\rangle_{b_{4}}\nonumber\\
-|V\rangle_{a_{3}}|V\rangle_{b_{3}}|V\rangle_{a_{4}}|V\rangle_{b_{4}}),
\end{eqnarray}
with the probability of $4|\alpha\beta|^{2}F(1-F)$.
Finally, by measuring the photons in $a_{4}b_{4}$ modes in $|\pm\rangle$ basis, they will get
a new mixed state
\begin{eqnarray}
\rho'_{a_{3}b_{3}}&=&[F^{2}+(1-F)^{2}]|\phi^{+}\rangle_{a_{3}b_{3}}\langle\phi^{+}|\nonumber\\
&+&2F(1-F)|\phi^{-}\rangle_{a_{3}b_{3}}\langle\phi^{-}|.\label{mixed5}
\end{eqnarray}
Here
\begin{eqnarray}
|\phi^{\pm}\rangle_{a_{3}b_{3}}=\frac{1}{\sqrt{2}}(|H\rangle_{a_{3}}|H\rangle_{b_{3}}\pm|V\rangle_{a_{3}}|V\rangle_{b_{3}}).
\end{eqnarray}
From Eq. (\ref{mixed5}), comparing with Eq. (\ref{mixed1}), the fidelity will essentially decrease if $F\in(\frac{1}{2}, 1)$.
Fortunately, this mixed state can be purified using the  conventional EPPs \cite{Pan1} in a next round. Generally speaking,
the phase-flip error cannot be purified directly, and it should be transformed to the bit-flip error with the Hadamard operation.
In the optical system, the quarter wave plate can act the role of the Hadamard operation and it makes
\begin{eqnarray}
|H\rangle\rightarrow\frac{1}{\sqrt{2}}(|H\rangle+|V\rangle),\nonumber\\
|V\rangle\rightarrow\frac{1}{\sqrt{2}}(|H\rangle-|V\rangle).
\end{eqnarray}
After this transformation, $\rho'_{a_{3}b_{3}}$ can be rewritten as
\begin{eqnarray}
\rho''_{a_{3}b_{3}}&=&[F^{2}+(1-F)^{2}]|\phi^{+}\rangle_{a_{3}b_{3}}\langle\phi^{+}|\nonumber\\
&+&2F(1-F)|\psi^{+}\rangle_{a_{3}b_{3}}\langle\psi^{-}|,\label{mixed6}
\end{eqnarray}
with
\begin{eqnarray}
|\psi^{+}\rangle_{a_{3}b_{3}}=\frac{1}{\sqrt{2}}(|H\rangle_{a_{3}}|V\rangle_{b_{3}}+|V\rangle_{a_{3}}|H\rangle_{b_{3}}).
\end{eqnarray}
By reusing the purification principle of  four-mode cases, the state of Eq. (\ref{mixed6}) can be purified into a  higher fidelity mixed state.
In this way, we can obtain a high quality entangled state from  arbitrary mixed states.

Actually, in the first step of the phase-flip distillation,
after they successfully  choosing the four-mode cases, they cannot perform the second step because the photons are destroyed by the photon detectors. In a practical
operation, they do not need to choose the four-mode cases if the single-photon detectors are available.
 From Eqs. (\ref{phase1})
and (\ref{phase2}),  the spatial modes $a_{4}$ and $b_{4}$ exactly containing one photon essentially means
the four-mode cases, because another two photons  are always in the spatial modes $a_{3}$ and $b_{3}$.

\section{Multipartite entanglement distillation}
It is straightforward to extend this protocol to the case of multi-partite system.
The multi-partite  Greenberger-Horne-Zeilinger (GHZ) state can be described as
\begin{eqnarray}
|\Phi^{+}\rangle_{N}=\frac{1}{\sqrt{2}}(|H\rangle|H\rangle\cdots |H\rangle+|V\rangle|V\rangle\cdots |V\rangle).\label{multipartite1}
\end{eqnarray}
\begin{figure}[!h]%[tpb]
\begin{center}
\includegraphics[width=9cm,angle=0]{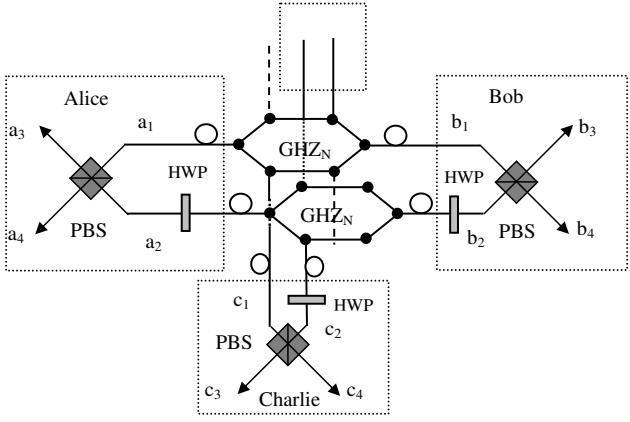}
\caption{A schematic drawing of our EDP for the multi-photon system. Each party say Alice, Bob and Charlie owns the same set up like Fig. 1.
By selecting all the output modes of the PBSs containing one photon, they can ultimately obtain a high quality N-photon entangled state.}
\end{center}
\end{figure}
The noisy channel will make the maximally entangled state $|\Phi^{+}\rangle_{N}$ become
\begin{eqnarray}
\rho_{N}=F|\Phi_{1}^{+}\rangle_{N}\langle\Phi_{1}^{+}|+(1-F)|\Psi_{1}^{+}\rangle_{N}\langle\Psi_{1}^{+}|.
\end{eqnarray}
Here
\begin{eqnarray}
|\Phi_{1}^{+}\rangle_{N}=\alpha|H\rangle|H\rangle\cdots |H\rangle+\beta|V\rangle|V\rangle\cdots |V\rangle,
\end{eqnarray}
and
\begin{eqnarray}
|\Psi_{1}^{+}\rangle_{N}=\alpha|V\rangle|H\rangle\cdots |H\rangle+\beta|H\rangle|V\rangle\cdots |V\rangle.
\end{eqnarray}
From Fig. 2, the mixed state shared by Alice, Bob, $\cdots$, in the spatial modes $a_{1}$, $b_{1}$, $\cdots$ is $\rho_{N}$.
The mixed state in the spatial modes  $a_{2}$, $b_{2}$, $\cdots$,  will first be flipped and becomes $\rho'_{N}$, with
\begin{eqnarray}
\rho'_{N}=F|\Phi_{1}^{+}\rangle^{\perp}_{N}\langle\Phi_{1}^{+}|^{\perp}+(1-F)|\Psi_{1}^{+}\rangle^{\perp}_{N}\langle\Psi_{1}^{+}|^{\perp}.
\end{eqnarray}
Here
\begin{eqnarray}
|\Phi_{1}^{+}\rangle^{\perp}_{N}=\alpha|V\rangle|V\rangle\cdots |V\rangle+\beta|H\rangle|H\rangle\cdots |H\rangle,
\end{eqnarray}
and
\begin{eqnarray}
|\Psi_{1}^{+}\rangle^{\bot}_{N}=\alpha|H\rangle|V\rangle\cdots |V\rangle+\beta|V\rangle|H\rangle\cdots |H\rangle.
\end{eqnarray}
In this way, the two pairs $\rho_{N}\otimes\rho'_{N}$ can be seen as the probabilistic
mixture of four pure states: with a probability of $F^{2}$, pairs are in the state
$|\Phi_{1}^{+}\rangle_{N}\otimes|\Phi_{1}^{+}\rangle^{\perp}_{N}$, with equal probability
of $F(1-F)$, in the states $|\Phi_{1}^{+}\rangle_{N}\otimes|\Psi_{1}^{+}\rangle^{\perp}_{N}$ and
$|\Psi_{1}^{+}\rangle_{N}\otimes|\Phi_{1}^{+}\rangle^{\perp}_{N}$, and with a probability of $(1-F)^{2}$
in the state $|\Psi_{1}^{+}\rangle_{N}\otimes|\Psi_{1}^{+}\rangle^{\perp}_{N}$.

Interestingly, the cross-combinations $|\Phi_{1}^{+}\rangle_{N}\otimes|\Psi_{1}^{+}\rangle^{\perp}_{N}$ and
$|\Psi_{1}^{+}\rangle_{N}\otimes|\Phi_{1}^{+}\rangle^{\perp}_{N}$ cannot lead all the output modes exactly
contain only one photon after they passing through the PBSs. They can be eliminated automatically.
Therefore, if they select the items which make all the output modes of the PBSs exactly contain one photon, they will get a $2N$-photon
mixed state.
That is, with the  probability of $2|\alpha\beta|^{2}F^{2}$, they are in the state
\begin{eqnarray}
|\Phi^{+}\rangle_{2N}&=&\frac{1}{\sqrt{2}}(|H\rangle_{a_{1}}|H\rangle_{b_{1}}\cdots |H\rangle_{a_{2}}\cdots |H\rangle_{k_{2}}\nonumber\\
&+&|V\rangle_{a_{1}}|V\rangle_{b_{1}}\cdots |V\rangle_{a_{2}}\cdots |V\rangle_{k_{2}}),\label{multi1}
\end{eqnarray}
 and with the probability of $2|\alpha\beta|^{2}(1-F)^{2}$, they are in the state
 \begin{eqnarray}
 |\Psi^{+}\rangle_{2N}&=&\frac{1}{\sqrt{2}}(|V\rangle_{a_{1}}|H\rangle_{b_{1}}\cdots |V\rangle_{a_{2}}|H\rangle_{b_{2}}\cdots |H\rangle_{k_{2}}\nonumber\\
&+&|H\rangle_{a_{1}}|V\rangle_{b_{1}}\cdots |H\rangle_{a_{2}}\cdots |V\rangle_{k_{2}}).\label{multi2}
 \end{eqnarray}
After  measuring the photons in the $|\pm\rangle$ basis, Eq. (\ref{multi1}) will become
 \begin{eqnarray}
 &&|\Phi^{+}\rangle'_{2N}=\frac{1}{\sqrt{2}}(|H\rangle_{a_{1}}|H\rangle_{b_{1}}\cdots|H\rangle_{k_{1}}\nonumber\\
 &&(\frac{1}{\sqrt{2}})^{\otimes N}(|+\rangle+|-\rangle)^{\otimes N}\nonumber\\
 &+&(|V\rangle_{a_{1}}|V\rangle_{b_{1}}\cdots|V\rangle_{k_{1}}(\frac{1}{\sqrt{2}})^{\otimes N}(|+\rangle-|-\rangle)^{\otimes N}.
 \end{eqnarray}
 The Eq. (\ref{multi2}) will become
  \begin{eqnarray}
  &&|\Psi^{+}\rangle'_{2N}=\frac{1}{\sqrt{2}}(|V\rangle_{a_{1}}|H\rangle_{b_{1}}\cdots|H\rangle_{k_{1}})\nonumber\\
  &&(\frac{1}{\sqrt{2}})^{\otimes N}(|+\rangle-|-\rangle)(|+\rangle+|-\rangle)^{\otimes N-1}\nonumber\\
  &&+|H\rangle_{a_{1}}|V\rangle_{b_{1}}\cdots |V\rangle_{k_{1}}(\frac{1}{\sqrt{2}})^{\otimes N}\nonumber\\
  &&(|+\rangle+|-\rangle)(|+\rangle-|-\rangle)^{\otimes N-1}.
  \end{eqnarray}
  Therefore, if the measurement results on the photons in the spatial modes $a_{2}$, $b_{2}$, $\cdots$ are the even number of $|+\rangle$, they will get the mixed state shown in Eq.(\ref{multipartite1}), and the fidelity is also $\frac{F^{2}}{F^{2}+(1-F)^{2}}$. Otherwise, they will get
 \begin{eqnarray}
|\Phi^{-}\rangle_{N}=\frac{1}{\sqrt{2}}(|H\rangle|H\rangle\cdots |H\rangle-|V\rangle|V\rangle\cdots |V\rangle),
\end{eqnarray}
with the same fidelity. In order to get the $|\Phi^{+}\rangle_{N}$, one of them should perform a phase-flip
on her or his photon.
Once the bit-flip error can be distilled successfully, with
 the same principle described in Sec. III, if a phase-flip error occurs, it can also be distilled successfully.

\section{Distillation protocol using SPDC source}
So far, we have fully explained this EDP theoretically. From the above discussion, we resort to the ideal sources
to realize this protocol. However, in current experimental technology, the ideal sources are not available.
In this section, we will discuss the experimental  realization of our EDP with current available SPDC source. We will show that with the SPDC source,
we can also achieve this EDP effectively.

The SPDC source generates the entanglement state of the form \cite{Yamamoto1}
\begin{eqnarray}
|\Upsilon\rangle=\sqrt{g}(|vac\rangle+p|\phi^{+}\rangle+p^{2}|\phi^{+}\rangle^{2}+\cdots).\label{SPDC}
\end{eqnarray}
Here $|vac\rangle$ is the vacuum state. $p$ is the probability for generating the $|\phi^{+}\rangle$. For a optical system, we
denote $|0\rangle\equiv|H\rangle$ and $|1\rangle\equiv|V\rangle$ and let $|\phi^{+}\rangle=\frac{1}{\sqrt{2}}(|H\rangle|H\rangle+|V\rangle|V\rangle)$. The $\sqrt{g}$ is a global phase factor which
can be omitted.
The feature of such SPDC source makes it  unsuitable for achieving the distillation task. The ideal source should emit exactly one pair of entangled states
in a given moment. However, if both the sources S$_{1}$ and S$_{2}$ are the SPDC sources, one cannot ensure that
the four photons come from both sources, for they may come from the same sources. It will lead this distillation protocol fail under current available experimental techniques. Interestingly,  we will show that
such a  SPDC source is not only suitable for distillation, but also can be more efficient than the original
ideal source, which will become an advantage of this protocol.

From Fig. 3, a pump pulse of
ultraviolet light passes through a beta barium borate (BBO)
crystal and produces correlated pairs of photons into the
modes $a_{1}$ and $b_{1}$. Then the pump pulse of ultraviolet light is reflected and traverses the crystal for
a second time, and produces correlated pairs of photons in
the modes $a_{2}$ and $b_{2}$.
The whole system can be described as
\begin{eqnarray}
&&|\Upsilon\rangle_{1}\otimes|\Upsilon\rangle_{2}\nonumber\\
&&=[\sqrt{g}(|vac\rangle+p|\phi^{+}\rangle_{a_{1}b_{1}}+p^{2}|\phi^{+}\rangle_{a_{1}b_{1}}^{2}+\cdots)]\nonumber\\
&\otimes&e^{i\Delta}[\sqrt{g}(|vac\rangle+p|\phi^{+}\rangle_{a_{2}b_{2}}+p^{2}|\phi^{+}\rangle_{a_{2}b_{2}}^{2}+\cdots)]\nonumber\\
&=&g[e^{i\Delta}|vac\rangle|vac\rangle+p(|\phi^{+}\rangle_{a_{1}b_{1}}+e^{i\Delta}|\phi^{+}\rangle_{a_{2}b_{2}})]\nonumber\\
&+&p^{2}[e^{i\Delta}|\phi^{+}\rangle_{a_{1}b_{1}}\otimes|\phi^{+}\rangle_{a_{2}b_{2}}+e^{i\Delta}|\phi^{+}\rangle^{2}_{a_{1}b_{1}}+|\phi^{+}\rangle^{2}_{a_{2}b_{2}}]\nonumber\\
&+&\cdots.\label{SPDC1}
\end{eqnarray}
Here the $\Delta$ is the relative phase between these two possibilities. It can become 0 by adjusting the relative position of the reflection
mirror $M$.  From Eq. (\ref{SPDC1}), the first term is $|vac\rangle|vac\rangle$ which means none photon.
With the probability of $p$, it is in the state $(|\phi^{+}\rangle_{a_{1}b_{1}}+e^{i\Delta}|\phi^{+}\rangle_{a_{2}b_{2}})$.
It is essentially the two-photon hyperentangled  state in the degrees of freedom of both polarization and spatial mode\cite{Simon,shengpra2,shengpra3,lixhepp}.
The four-photon state is $e^{i\Delta}|\phi^{+}\rangle_{a_{1}b_{1}}\otimes|\phi^{+}\rangle_{a_{2}b_{2}}+e^{i\Delta}|\phi^{+}\rangle^{2}_{a_{1}b_{1}}+|\phi^{+}\rangle^{2}_{a_{2}b_{2}}$,
with the probability of $p^{2}$.  The first part $|\phi^{+}\rangle_{a_{1}b_{1}}\otimes|\phi^{+}\rangle_{a_{2}b_{2}}$ denotes
the ideal source discussed in Sec. II. That is the spatial modes $a_{1}b_{1}$ and $a_{2}b_{2}$ exactly contain
one photon pair.  The $|\phi^{+}\rangle^{2}_{a_{1}b_{1}}$ and $|\phi^{+}\rangle^{2}_{a_{2}b_{2}}$
mean that the two photon pairs are in the same spatial modes. If we choose the four-mode cases, obviously, the
vacuum state and the two-photon state will have no contribution  to the distillation.
Interestingly, the two pairs in the same modes will also increase the fidelity
of the mixed state.

\begin{figure}[!h]%[tpb]
\begin{center}
\includegraphics[width=8cm,angle=0]{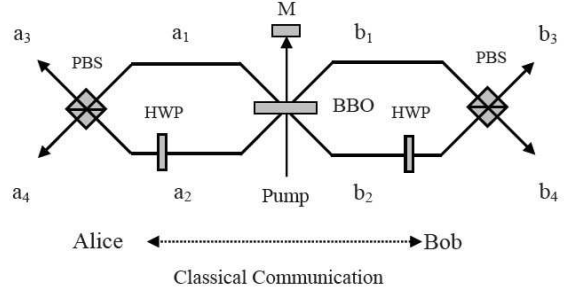}
\caption{Our distillation protocol with current SPDC source. The SPDC source can produce one photon pair with the probability
of $p$ and produce two photon pairs with the probability of $p^{2}$. Surprisingly,  under this condition, our protocol becomes more
efficient than that described for using the ideal sources.}
\end{center}
\end{figure}

In a practical experiment, we need to process two photon pairs which are in the mixed state in Eq. (\ref{mixed1}).
We also require to interfere the two photons at each sides at a PBS by making them indistinguishable.
In Refs. \cite{Pan2,zhao2,Yamamoto2}, they have detailed the preparation of the standard mixed states and the pure less-entangled states.
For example, to prepare the mixed state they send each pair through a half-wave plate whose angle is randomly set at either $+\delta$ or $-\delta$ \cite{Pan2}. They choose $\delta=14^{\circ}$ to make $F=0.75$. To prepare the less-entangled state they let the photon pair pass through the Brewster's windows
with the vertical axis tilted at $\theta$. In Ref. \cite{zhao2}, they choose $\theta=56^{\circ}$ which
makes $T_{H}=0.98$ for horizontal polarization photon and $T_{V}=0.73$ for vertical polarization photon.
Therefore, one can prepare the mixed state in Eq. (\ref{mixed1}) by combining such operations . The first term  $|\phi^{+}\rangle_{a_{1}b_{1}}\otimes|\phi^{+}\rangle_{a_{2}b_{2}}$ essentially denote the ideal source which has been discussed in Sec. II and we should not re-discuss any more. We only need to
discuss the contribution of  two pairs being in the same spatial mode.
If the two-photon pairs are both in the $a_{1}b_{1}$ modes, they will become
\begin{eqnarray}
&&|\Phi^{+}\rangle_{a_{1}b_{1}}\otimes|\Phi^{+}\rangle_{a_{1}b_{1}}=(\alpha|H\rangle_{a_{1}}|H\rangle_{b_{1}}+\beta|V\rangle_{a_{1}}|V\rangle_{b_{1}})\nonumber\\
&\otimes&(\alpha|H\rangle_{a_{1}}|H\rangle_{b_{1}}+\beta|V\rangle_{a_{1}}|V\rangle_{b_{1}})\nonumber\\
&=&\alpha^{2}|H\rangle_{a_{1}}|H\rangle_{a_{1}}|H\rangle_{b_{1}}|H\rangle_{b_{1}}+\beta^{2}|V\rangle_{a_{1}}|V\rangle_{a_{1}}|V\rangle_{b_{1}}|V\rangle_{b_{1}}\nonumber\\
&+&\alpha\beta(|H\rangle_{a_{1}}|V\rangle_{a_{1}}|H\rangle_{b_{1}}|V\rangle_{b_{1}}+|V\rangle_{a_{1}}|H\rangle_{a_{1}}|V\rangle_{b_{1}}|H\rangle_{b_{1}})\nonumber\\
&\rightarrow&\alpha^{2}|H\rangle_{a_{4}}|H\rangle_{a_{4}}|H\rangle_{b_{4}}|H\rangle_{b_{4}}+\beta^{2}|V\rangle_{a_{3}}|V\rangle_{a_{3}}|V\rangle_{b_{3}}|V\rangle_{b_{3}}\nonumber\\
&+&2\alpha\beta|V\rangle_{a_{3}}|V\rangle_{b_{3}}|H\rangle_{a_{4}}|H\rangle_{b_{4}},\label{spdcremain1}
\end{eqnarray}
with the probability of $F$, and they will become
\begin{eqnarray}
&&|\Psi^{+}\rangle_{a_{1}b_{1}}\otimes|\Psi^{+}\rangle_{a_{1}b_{1}}=(\alpha|H\rangle_{a_{1}}|V\rangle_{b_{1}}+\beta|V\rangle_{a_{1}}|H\rangle_{b_{1}})\nonumber\\
&\otimes&(\alpha|H\rangle_{a_{1}}|V\rangle_{b_{1}}+\beta|V\rangle_{a_{1}}|H\rangle_{b_{1}})\nonumber\\
&=&\alpha^{2}|H\rangle_{a_{1}}|H\rangle_{a_{1}}|V\rangle_{b_{1}}|V\rangle_{b_{1}}+\beta^{2}|V\rangle_{a_{1}}|V\rangle_{a_{1}}|H\rangle_{b_{1}}|H\rangle_{b_{1}}\nonumber\\
&+&\alpha\beta(|H\rangle_{a_{1}}|V\rangle_{a_{1}}|V\rangle_{b_{1}}|H\rangle_{b_{1}}+|V\rangle_{a_{1}}|H\rangle_{a_{1}}|H\rangle_{b_{1}}|V\rangle_{b_{1}}).\nonumber\\
&\rightarrow&\alpha^{2}|H\rangle_{a_{4}}|H\rangle_{a_{4}}|V\rangle_{b_{3}}|V\rangle_{b_{3}}+\beta^{2}|V\rangle_{a_{3}}|V\rangle_{a_{3}}|H\rangle_{b_{4}}|H\rangle_{b_{4}}\nonumber\\
&+&2\alpha\beta|V\rangle_{a_{3}}|V\rangle_{b_{3}}|H\rangle_{a_{4}}|H\rangle_{b_{4}},\label{spdcremain2}
\end{eqnarray}
with the probability of $1-F$. We should point out that the cross-combination items $|\Phi^{+}\rangle_{a_{1}b_{1}}\otimes|\Psi^{+}\rangle_{a_{1}b_{1}}$ does not appear because the two pairs are operated simultaneously. They are either $|\Phi^{+}\rangle_{a_{1}b_{1}}$ or $|\Psi^{+}\rangle_{a_{1}b_{1}}$.
On the other hand, if the two-photon pairs are in the spatial mode $a_{2}b_{2}$, they will become
\begin{eqnarray}
&&|\Phi^{+}\rangle^{\perp}_{a_{2}b_{2}}\otimes|\Phi^{+}\rangle^{\perp}_{a_{2}b_{2}}=(\alpha|V\rangle_{a_{2}}|V\rangle_{b_{2}}+\beta|H\rangle_{a_{2}}|H\rangle_{b_{2}})\nonumber\\
&\otimes&(\alpha|V\rangle_{a_{2}}|V\rangle_{b_{2}}+\beta|H\rangle_{a_{2}}|H\rangle_{b_{2}})\nonumber\\
&=&\alpha^{2}|V\rangle_{a_{2}}|V\rangle_{a_{2}}|V\rangle_{b_{2}}|V\rangle_{b_{2}}+\beta^{2}|H\rangle_{a_{2}}|H\rangle_{a_{2}}|H\rangle_{b_{2}}|H\rangle_{b_{2}}\nonumber\\
&+&\alpha\beta(|H\rangle_{a_{2}}|V\rangle_{a_{2}}|H\rangle_{b_{2}}|V\rangle_{b_{21}}+|V\rangle_{a_{2}}|H\rangle_{a_{2}}|V\rangle_{b_{2}}|H\rangle_{b_{2}})\nonumber\\
&\rightarrow&\alpha^{2}|V\rangle_{a_{4}}|V\rangle_{a_{4}}|V\rangle_{b_{4}}|V\rangle_{b_{4}}+\beta^{2}|H\rangle_{a_{3}}|H\rangle_{a_{3}}|H\rangle_{b_{3}}|H\rangle_{b_{3}}\nonumber\\
&+&2\alpha\beta|H\rangle_{a_{3}}|H\rangle_{b_{3}}|V\rangle_{a_{4}}|V\rangle_{b_{4}},\label{spdcremain3}
\end{eqnarray}
with a probability of $F$, and
\begin{eqnarray}
&&|\Psi^{+}\rangle^{\perp}_{a_{2}b_{2}}\otimes|\Psi^{+}\rangle^{\perp}_{a_{2}b_{2}}=(\alpha|V\rangle_{a_{2}}|H\rangle_{b_{2}}+\beta|H\rangle_{a_{2}}|V\rangle_{b_{2}})\nonumber\\
&\otimes&(\alpha|V\rangle_{a_{2}}|H\rangle_{b_{2}}+\beta|H\rangle_{a_{2}}|V\rangle_{b_{2}})\nonumber\\
&=&\alpha^{2}|V\rangle_{a_{2}}|V\rangle_{a_{2}}|H\rangle_{b_{2}}|H\rangle_{b_{2}}+\beta^{2}|H\rangle_{a_{2}}|H\rangle_{a_{2}}|V\rangle_{b_{2}}|V\rangle_{b_{2}}\nonumber\\
&+&\alpha\beta(|H\rangle_{a_{2}}|V\rangle_{a_{2}}|H\rangle_{b_{2}}|V\rangle_{b_{21}}+|V\rangle_{a_{2}}|H\rangle_{a_{2}}|V\rangle_{b_{2}}|H\rangle_{b_{2}})\nonumber\\
&\rightarrow&\alpha^{2}|V\rangle_{a_{3}}|V\rangle_{a_{3}}|H\rangle_{b_{3}}|H\rangle_{b_{3}}+\beta^{2}|H\rangle_{a_{4}}|H\rangle_{a_{4}}|V\rangle_{b_{4}}|V\rangle_{b_{4}}\nonumber\\
&+&2\alpha\beta|H\rangle_{a_{3}}|H\rangle_{b_{3}}|V\rangle_{a_{4}}|V\rangle_{b_{4}},\label{spdcremain4}
\end{eqnarray}
with a probability of $1-F$. As discussed in Eq. (\ref{SPDC1}),  the two photon pairs in the spatial modes $a_{1}b_{1}$
and $a_{2}b_{2}$ will have a fixed relative phase $\Delta$, which can be adjusted to 0. After choosing the four-mode cases, such states thus are in a coherent superposition \cite{Pan2}
\begin{eqnarray}
\frac{1}{\sqrt{2}}(|V\rangle_{a_{3}}|V\rangle_{b_{3}}|H\rangle_{a_{4}}|H\rangle_{b_{4}}+e^{i\Delta}|H\rangle_{a_{3}}|H\rangle_{b_{3}}|V\rangle_{a_{4}}|V\rangle_{b_{4}}).\nonumber\\
\end{eqnarray}

\begin{figure}[!h]%[tpb]
\begin{center}
\includegraphics[width=7cm,angle=0]{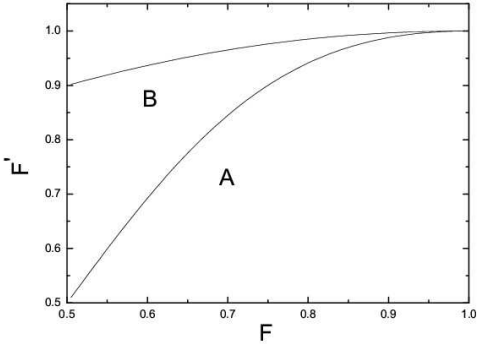}
\caption{The fidelity $F'$ of the purified  mixed state is altered by
the initial fidelity $F$ by performing the protocol one time. Curve A represents our protocol with ideal source.
Curve B is the fidelity of Eq. (\ref{fidelity}), which represents our protocol with the SPDC source. Obviously, with SPDC source, the fidelity increases
rapidly because of the contribution of two photon pairs being in the same spatial modes.}
\end{center}
\end{figure}

\begin{figure}[!h]%[tpb]
\begin{center}
\includegraphics[width=7cm,angle=0]{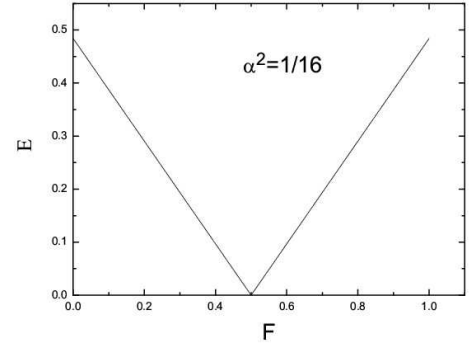}
\caption{Entanglement E is altered by
the initial fidelity $F$. Here we choose $\alpha^{2}=\frac{1}{16}$. It is shown that
the max value cannot reach 1.}
\end{center}
\end{figure}

 After we performing the polarization measurement
in the spatial modes $a_{4}b_{4}$ in the $|\pm\rangle$ basis, above state will also contribute to a
maximally entangled state $|\phi^{+}\rangle_{a_{3}b_{3}}$, which is exactly the desired maximally entangled state.
Our analysis is essentially analogy  with the experiment results reported by Ref. \cite{Pan2}, and we show that
in the practical experiment, the SPDC source is also suitable for current distillation protocol. Moreover, the two photon pairs in the
same spatial modes will contribute to the additional maximally entangled pair to increase the whole fidelity. We can recalculate the fidelity
by using the SPDC source, with
\begin{eqnarray}
F'&=&\frac{2|\alpha\beta|^{2}F^{2}+4|\alpha\beta|^{2}}{2|\alpha\beta|^{2}F^{2}+2|\alpha\beta|^{2}(1-F)^{2}+4|\alpha\beta|^{2}}\nonumber\\
&=&\frac{F^{2}+2}{F^{2}+(1-F)^{2}+2}.\label{fidelity}
\end{eqnarray}

In Fig. 4, we calculate the fidelity after performing the protocol one time. We choose the initial fidelity $F\in(0.5,1)$. It is shown
that using SPDC source, the fidelity increases rapidly.

\section{discussion and conclusion}

By far, we have briefly explained this  EDP. It is
interesting to compare this protocol with PBS-purification protocol \cite{Pan1}. In PBS-purification protocol, they used the
PBSs to substitute the CNOT gates and chose the four-mode cases to achieve the purification.
That is to say, the PBS essentially acts the similar role as the CNOT gate. Because it is an implementation
of the CNOT gate between a spatial mode qubit and a polarization qubit. The spatial mode is flipped
or not flipped as a function of the polarization. On the other hand, in a standard entanglement concentration
protocol, the PBS can also act the same role as the CNOT gate. In this protocol, it is essentially the combination of the
entanglement purification and entanglement concentration. Interestingly, this distillation task can be achieved simultaneously, by selecting
the four-mode cases. In PBS-purification protocol, the success probability is $F^{2}+(1-F)^{2}$, while in PBS-concentration protocols, the
success probability is $2|\alpha\beta|^{2}$\cite{Pan1,zhao1}. In this protocol, it is $2|\alpha\beta|^{2}[F^{2}+(1-F)^{2}]$. It is not only decided by the initial
fidelity $F$, but also decided by the initial coefficient $\alpha$ and $\beta$. One can find that if $\alpha=\beta=\frac{1}{\sqrt{2}}$, this distillation
protocol is equal to the PBS-purification protocol \cite{Pan1}. On the other hand, if the original state is a pure less-entangled state with $F=1$, our distillation
protocol essentially is a standard EPP as the same as Ref. \cite{zhao1}.

In the conventional EPP, the phase-flip error cannot be purified
directly. It  is usually converted into a bit-flip error and  purified in the second round.
In this way, the entanglement purification with CNOT gates is consistent with the protocol  with PBSs.
However, if we purify the phase-flip error directly both using the CNOT gate and PBS gate, we find that
they are different \cite{C.H.Bennett1,Pan1}. For exmaple, in Ref. \cite{C.H.Bennett1}, if the initial mixed state is $\rho=F|\phi^{+}\rangle\langle\phi^{+}|+(1-F)|\phi^{-}\rangle\langle\phi^{-}|$, after performing the EPP, the $F$ does not change. However, in PBS-purification protocol,
after performing the four-mode cases, the fidelity is  $F'=F^{2}+(1-F)^{2}<F$, if $F\in(0.5,1)$.
Certainly, it does not affect the purification of phase-flip error because it is always  flipped
into a bit-flip error. However, in this protocol, we cannot treat the phase-flip error like this.
 In Eq. (\ref{mixed3}),  with the probability of $F$ it is in $|\Phi^{+}\rangle_{ab}$, and with a probability
of $1-F$, it is in the state $|\Phi^{-}\rangle_{ab}$. Both of them are still the  pure less-entangled states.
Different from the Eq. (\ref{mixed1}), we cannot adopt the conventional way to convert the phase-flip error
into the bit-flip error directly. We should concentrate the state in Eq. (\ref{mixed3})
 into a mixed state with each items being the maximally entangled state shown in Eq. (\ref{mixed5}).
In this process, we find that the fidelity decreases. It is not strange that the fidelity's decreasing.
Because both concentration and purification are based on the LOCC.
LOCC  cannot increase the entanglement. That is, the entanglement after performing the LOCC operations should
be less than or equal to the initial entanglement. The entanglement distillation is essentially
the transformation of entanglement. In Eq. (\ref{mixed3}), each items are less-entangled state but in
Eq. (\ref{mixed5}) each items are the maximally entangled state. In this way, the fidelity of Eq. (\ref{mixed5})
must be lower than it is in Eq. (\ref{mixed3}) to ensure the total entanglement does not increase.

\begin{figure}[!h]%[tpb]
\begin{center}
\includegraphics[width=7cm,angle=0]{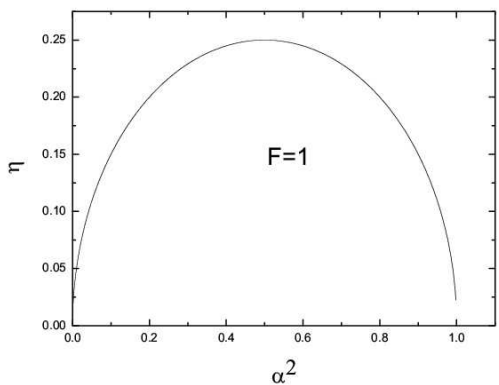}
\caption{Entanglement transformation efficiency $\eta$ is altered by
the initial coefficient  $\alpha^{2}$. Here we choose $F=1$ which is a linear optical ECP shown in Ref. \cite{zhao1}. This result agrees with
the Ref. \cite{shengsinglephotonconcentration}.}
\end{center}
\end{figure}

\begin{figure}[!h]%[tpb]
\begin{center}
\includegraphics[width=7cm,angle=0]{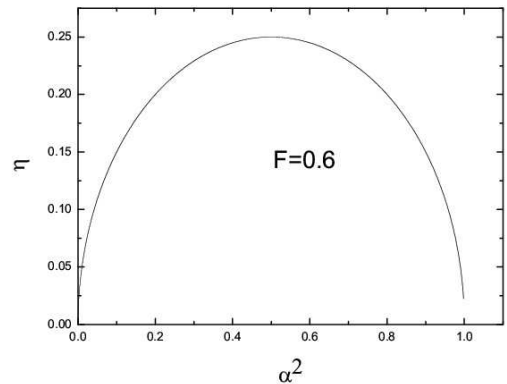}
\caption{Entanglement transformation efficiency $\eta$ is altered by
the initial coefficient  $\alpha^{2}$. Here we choose $F=0.6$. Interestingly, compared with Fig. 6, our numerical simulation
results show that the  $\eta$ does not change with the initial fidelity $F$ but it changes with the coefficient $\alpha$.}
\end{center}
\end{figure}

 Following the same principle of Ref. \cite{shengsinglephotonconcentration},
we can denote the efficiency of such entanglement transformation like
 \begin{eqnarray}
 \eta=\frac{E'_{d}*P}{2E^{0}_{d}},
  \end{eqnarray}
  in this distillation protocol.
The subscription $d$ means this distillation protocol. The $E'_{d}*P$ means the entanglement of the remaind mixed state
with the success probability of $P$. If we take the bit-flip error for an example, one can find that $P=2|\alpha\beta|^{2}(F^{2}+(1-F)^{2})$.
 The $2E^{0}_{d}$ means that before performing this protocol, we need two pairs of
original entangled states. For a two-qubit system, the concurrence is a good  indicator for measuring the entanglement \cite{concurrence}.
We first calculate the entanglement $E$ which is changed with the initial fidelity $F$ and the coefficient
$\alpha$. Fig. 5 shows the relationship between the $E$ and $F$ with $\alpha^{2}$ being a constant.
In Fig. 5,
we choose $\alpha^{2}=\frac{1}{16}$. For a mixed state showing in Eq. (\ref{mixed1}), the entanglement is equal to 0 when
$F=0.5$. Interestingly, our results show that it is a linear relationship between $\eta$  and $F$ if $\alpha^{2}$ is a constant.
The difference is that the max value of $\eta$ is lower than 1 when $\alpha^{2}\neq\frac{1}{2}$.
We also calculate
the relationship between $\eta$ and $\alpha^{2}$ with $F$ being a constant. In Fig. 6, we choose $F=1$ and it is correspond
to the standard ECP \cite{zhao1}. Our numerical simulation results is consistent with Ref. \cite{shengsinglephotonconcentration},
which describe the entanglement with von Neumann entropy. Fig. 7 shows the relationship between $\eta$ and $\alpha^{2}$ when
$F=0.6$.

In conclusion, we have present a practical EDP for the general mixed state.
It contains both the purification and concentration. We both discussed the cases of distillation of a bit-flip error and a phase-flip error.
  If a bit-flip error occurs, the purification and
concentration procedure  can be completed simultaneously. If a phase-flip error occurs, we should first
perform the concentration procedure and perform the purification next. This protocol is also suitable for
the multi-partite system. We also discussed the experimental realization in current technology using available
SPDC source.  With SPDC source, after performing this protocol, the fidelity increases rapidly than using
the ideal source. We hope that this EDP is useful in practical long-distance quantum communications.

\section*{ACKNOWLEDGEMENTS}
We thank Dr. Bao-Kui Zhao (Jilin University, China) for
helpful discussion. This work is supported by the National Natural Science Foundation of
China under Grant No. 11104159  and 11347110,
and the Project
Funded by the Priority Academic Program Development of Jiangsu
Higher Education Institutions.

\end{document}